# Neutrino spin oscillations in gravitational fields in noncommutative Spaces

## S. A. Alavi;  S. Nodeh

*Department of Physics, Hakim Sabzevari University, P. O. Box 397, Sabzevar, Iran.*

*s.alavi@hsu.ac.ir; alaviag@gmail.com*

*We study neutrino spin oscillations in gravitational fields in non-commutative spaces. For the Schwarzschild  metric the maximum frequency decreases with increasing the noncommutativity parameter. In the case of  Reissner-Nordstrom (RN) metric, the maximum frequency of oscillation is a monotonically increasing function of  the  noncommutativity  parameter .In  both cases, the frequency of spin oscillations decreases as the distance from the  gravitational source  grows. We present a phenomenological application of our results. It is also shown that the  noncommutativity  parameter is bounded as $\sqrt{\theta} > 0.1\, l_p$.*

## Introduction

As it is well  known  gravitation  is  one of  the  most  studied phenomena in physics. The investigation  during  the  last century have led to important insights. Interaction of particles with gravitational field  would allow us to probe the   physics at the  Planck  scale  and increase our knowledge  about the underlying yet to be found theory. It  can also teach us to find the  missing link between general relativity and quantum theory.

Recently there  have  been  notable  studies  on  the  formulation  and possible experimental consequences of extensions of theories in the non-commutative spaces. There has been also a growing interest in possible cosmological consequences of  space  noncommutativity, see e.g., [1].

For a manifold parametrized by the coordinates $x_i$, the non-commutative relations can be written as: $[\hat{x}_i, \hat{x}_j] = i\theta_{ij}$, $[\hat{x}_i, \hat{P}_j] = i\delta_{ij}$, $[\hat{P}_i, \hat{P}_j] = 0$, where $\theta_{ij}$ is an antisymmetric tensor, see [2] for a review. Coordinate noncommutativity implies the existence of a finite minimal length $\sqrt{\theta}$, below which the concept of "distance" becomes physically meaningless so in non-commutative spaces there can't be point like objects. To study gravitational field in the non-commutative spaces, it is not necessary to change the Einstein's tensor part of the field equations and non-commutative effects act only on the matter source. So one can modify the distribution of point like source in favor of smeared objects. Noncommutativity replaces point-like structures by smeared objects and so may eliminate the divergences that normally appear in general relativity. The effect of smearing is mathematically implemented as follows: position Dirac-delta function is replaced everywhere with a Gaussian distribution of minimum width $\sqrt{\theta}$. So we choose the mass density of a static, spherically symmetric, smeared, particle- like gravitational source as [3,4] :

$$\rho_\theta(r) = \frac{M}{(4\pi\theta)^{\frac{3}{2}}} \exp\left(-\frac{r^2}{4\theta}\right). \tag{1}$$

A particle of mass M, instead of being perfectly localized at a point is diffused throughout a region of line size $\sqrt{\theta}$.

On the other hand the neutrino physics is an active area of research with important implications for particle physics, cosmology and astrophysics [5]. The physics associated with the neutrino oscillation is an interesting subject to study from both theoretical and experimental point of view. The phenomenon of neutrino oscillations can explain solar and atmospheric neutrino problems. It also provides the first experimental evidence for physics beyond the standard model since it requires nonzero mass for neutrinos. Interaction of neutrinos with an external field provides one of the factors required for a transition between helicity states. In this paper, we study neutrino spin oscillations in Schwarzschild and RN metrics in non-commutative spaces. We recall that the effect of a quantum gravity-induced minimal length on neutrino oscillations has been studied in [6].

# Neutrino spin oscillations in noncommutative Schwarzschild metric

By solving the Einstein's equations with $\rho_\theta(r)$ as a matter source and setting $\hbar = c = 1$, we have [7]:

$$ds^2 = \left(1 - \frac{4M}{r\sqrt{\pi}}\gamma\left(\frac{3}{2}, \frac{r^2}{4\theta}\right)\right)dt^2 - \left(1 - \frac{4M}{r\sqrt{\pi}}\gamma\left(\frac{3}{2}, \frac{r^2}{4\theta}\right)\right)^{-1}dr^2 - r^2\left(d\vartheta^2 + \sin^2\vartheta d\varphi^2\right),\tag{2}$$

where :

$$\gamma\left(\frac{3}{2}, \frac{r^2}{4\theta}\right) = \int_0^{\frac{r^2}{4\theta}} dt\, t^{1/2}e^{-t}.\tag{3}$$

is the lower incomplete Gamma function.

Here the mass M is diffused throughout a region of line size $\sqrt{\theta}$. The components of vierbein four velocity are as follows [8]:

$$u^a = \left(\alpha A, U_r A^{-1}, U_\theta r, U_\varphi r \sin\theta\right),\tag{4}$$

where

$$U^\mu = \left(U^0, U_r, U_\theta, U_\varphi\right),\ \alpha = U^0 = \frac{dt}{d\tau}.\tag{5}$$

is the four velocity of a particle in its geodesic path, which is related to vierbein four velocity through $u^a = e^a_\mu U^\mu$. The four velocity of a particle in the relevant metric $U^\mu$ is related to the world velocity of the particle through $\vec{U} = \alpha\vec{V}$ where $\alpha = \frac{dt}{d\tau}$ and $\tau$ is the proper time. The non-zero vierbein vectors are presented in the Appendix. To study the spin evolution of a particle in a gravitational field, we calculate $G_{ab} = \left(\vec{E}, \vec{B}\right)$ which is the analogue of the electromagnetic field tensor. It is defined as follows:

$$G_{ab} = e_{a\mu;\nu}e^\mu_b U^\nu.\tag{6}$$

where $e_{a\mu;\nu}$ are the covariant derivatives of vierbein vectors which are defined through the following expression:

$$e_{a\mu;\nu} = \frac{\partial e_{a\mu}}{\partial x^\nu} - \Gamma^\lambda_{\mu\nu}e_{a\lambda}.\tag{7}$$

Using this equation one can calculate the covariant derivatives of vierbein vectors (see the Appendix).

Using the fact that any anti-symmetric tensor in four-dimensional Minkowskian space-time can be stated in terms of two three dimensional vectors (such as electric and magnetic fields), we have:

$$G_{ab} = (\vec{E}, \vec{B}) \quad , G_{0i} = E_i \ , \ G_{ij} = -\epsilon_{ijk} B_k \tag{8}$$

Using Eqs. (5), (6) and (8) we have the following forms for the analogues of the electric and magnetic fields:

$$\vec{E} = \left( -\frac{1}{2} \left( \frac{4M}{r^2 \sqrt{\pi}} \frac{2}{3} X^{3/2} - \frac{4M}{r \sqrt{\pi}} X^{1/2} e^{-X} \right) u^t, 0, 0 \right),$$

$$\vec{B} = \left( \alpha \cos\vartheta \, v_\varphi, -\sin\vartheta \left( 1 - \frac{4M}{r\sqrt{\pi}} \frac{2}{3} X^{\frac{3}{2}} \right)^{\frac{1}{2}} \alpha v_\varphi, \left( 1 - \frac{4M}{r\sqrt{\pi}} \gamma \left( \frac{3}{2}, X \right) \right)^{\frac{1}{2}} \alpha v_\vartheta \, , \right), \tag{9}$$

where $\alpha = u^t = \frac{dt}{d\tau}$.

Geodesic equation of a particle in a gravitational field is as follows [9]:

$$\frac{d^2 x^\mu}{dq^2} + \Gamma^\mu_{\sigma\nu} \frac{dx^\sigma}{dq} \frac{dx^\nu}{dq} = 0 \ . \tag{10}$$

where the variable $q$ parameterizes the particle's world line. For a circular orbit with constant radius $r$ ($U_r = \frac{dr}{d\tau} = 0$). From Eqs. (2) and (10), we can calculate the values of $v_\varphi$ and $\alpha^{-1}$:

$$\alpha^{-1} = = \frac{d\tau}{dt} = \sqrt{A - r \frac{A'}{2}} \ , \tag{11}$$

$$v_\varphi = \sqrt{\frac{A'}{2r}} \ , \tag{12}$$

where $A' = \frac{d \left( 1 - \frac{4M}{r\sqrt{\pi}} \gamma \left( \frac{3}{2}, X \right) \right)^{\frac{1}{2}}}{dr}$.

Neutrino spin precession is given by the expression $\vec{\Omega} = \frac{\vec{G}}{\alpha}$ where vector $G$ is defined as follows [8]:

$$\vec{G} = \frac{1}{2}\left[\vec{B} + \frac{1}{1+u^0}[\vec{E} \times \vec{u}]\right]$$

(13)

By substituting (4), (9), (11) and (13) in $\overrightarrow{\Omega} = \frac{\vec{G}}{\alpha}$, the only non-zero component of frequency i.e. $\Omega_2$ is obtained:

$$\Omega_2 = \frac{v_\varphi}{2}\left[-\left(1 - \frac{4M}{r\sqrt{\pi}}\gamma\left(\frac{3}{2}, X\right)\right)^{\frac{1}{2}} + \frac{\alpha}{1+u^0}\left(1 - \frac{4M}{r\sqrt{\pi}}\gamma\left(\frac{3}{2}, X\right) - \frac{4M}{r\sqrt{\pi}}\frac{d}{dr}\gamma\left(\frac{3}{2}, X\right)\right)\right] = -\frac{v_\varphi}{2}\alpha^{-1}.$$

(14)

If $X \to \infty$ we recover the results of the commutative case [8]:

Using Eq. (14), we plot $2|\Omega_2|M$ versus $\frac{r}{2M}$ for different values of $\eta = \frac{\sqrt{\theta}}{M}$.

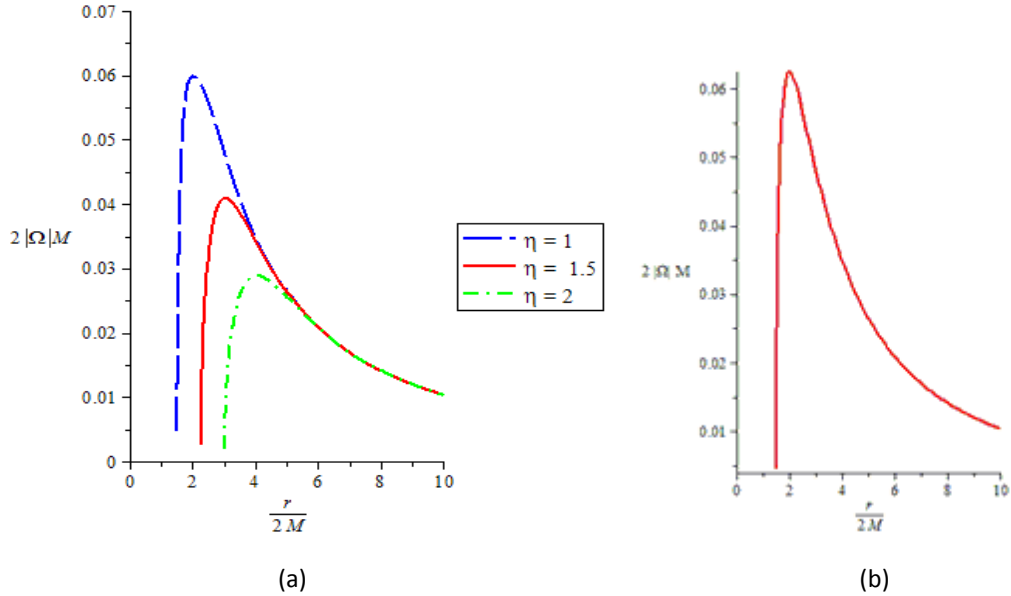

(a)                                                                (b)

Figure 1: Neutrino spin oscillations frequency versus radius of the neutrino orbit for different values of $\eta$ (a), and the commutative case $X \to \infty$ (b).

We observe that the maximum frequency decreases with increasing the noncommutativity parameter.

# Neutrino spin oscillations in noncommutative Reissner-Nordstrom (RN) metric.

In this section we investigate the effects of charge on neutrino spin oscillations in gravitational field, so we start with RN metric. In comparison with the previous calculations, it is interesting to study whether it produces some new results.

The RN metric in a non-commutative space is given by [10,11]:

$$ds^2 = g_{00}dt^2 - g_{00}^{-1}dr^2 - r^2 d\vartheta^2 - r^2\sin^2\vartheta d\varphi^2,$$

where:

$$g_{00} = A^2 = 1 - \frac{4M}{r\sqrt{\pi}}\gamma\left(\frac{3}{2}, X\right) + \frac{Q^2}{\pi r^2}\gamma^2\left(\frac{1}{2}, X\right) - \frac{Q^2}{\pi r\sqrt{2\theta}}\gamma\left(\frac{1}{2}, \frac{r^2}{2\theta}\right). \tag{15}$$

Here $Q$ and M are the charge and the mass of the gravitational source respectively.

The non-zero vierbein vectors and their covariant derivatives are presented in the Appendix.

Using Eqs.(5), (6) and (8) we have the following forms for the gravi-electric and gravi-magnetic fields:

$$E_1 = \frac{\alpha}{2} \cdot \left[ \begin{array}{c} \frac{4M}{r^2\sqrt{\pi}}\gamma\left(\frac{3}{2}, \frac{r^2}{4\theta}\right) - \frac{4M}{r\sqrt{\pi}}\frac{d}{dr}\gamma\left(\frac{3}{2}, \frac{r^2}{4\theta}\right) - \frac{2Q^2}{\pi r^3}\gamma\left(\frac{1}{2}, \frac{r^2}{4\theta}\right)^2 + \frac{Q^2}{\pi r^2}\frac{d}{dr}\gamma^2\left(\frac{1}{2}, \frac{r^2}{4\theta}\right) + \frac{Q^2}{\pi r^2\sqrt{2\theta}}\gamma\left(\frac{1}{2}, \frac{r^2}{2\theta}\right) \\ -\frac{Q^2}{\pi r\sqrt{2\theta}}\frac{d}{dr}\gamma\left(\frac{1}{2}, \frac{r^2}{2\theta}\right) \end{array} \right]$$

$$E_2 = 0, \quad E_3 = 0,$$

$$B_1 = -\alpha v_\varphi \cos\vartheta, \quad B_2 = \alpha v_\varphi A \sin\vartheta, \quad B_3 = \alpha A v_\theta. \tag{16}$$

From Eqs. (10) and (15) we can calculate the values of $v_\varphi$ and $\alpha^{-1}$:

$$\alpha^{-1} = \sqrt{A - r\frac{\acute{A}}{2}}, \quad v_\varphi = \sqrt{\frac{\acute{A}}{2r}}, \tag{17}$$

where $\acute{A} = \dfrac{d\left(1 - \frac{4M}{r\sqrt{\pi}}\gamma\left(\frac{3}{2}, X\right) + \frac{Q^2}{\pi r^2}\gamma^2\left(\frac{1}{2}, X\right) - \frac{Q^2}{\pi r\sqrt{2\theta}}\gamma\left(\frac{1}{2}, \frac{r^2}{2\theta}\right)\right)^{\frac{1}{2}}}{dr}.$

By substituting (4), (13), (16) and (17) in $\vec{\Omega} = \frac{\vec{G}}{\alpha}$, the only non-zero component of frequency i.e. $\Omega_2$ is obtained:

$$\Omega_2 = -\frac{1}{2}\sqrt{\frac{A'}{2r}}\sqrt{\left(A - r\frac{A'}{2}\right)} = -\frac{1}{2}\alpha^{-1}v_\varphi \ .$$

(18)

Using Eq. (18), we have  plotted  $2|\Omega_2|M$  versus  $\frac{r}{2M}$  for different values of  $\eta = \sqrt{\theta}/M$ and the commutative case in Fig.(2). It is seen that the maximum frequency of oscillation is a monotonically increasing function of the noncommutativity   parameter.

One can recover  the results  of  the  Schwarzschild  metric (Eq. (14))  as a special case  by taking $Q = 0$. It can be also shown that in the  limit  $X \to \infty$, we recover the results of  the commutative case  presented in [12].

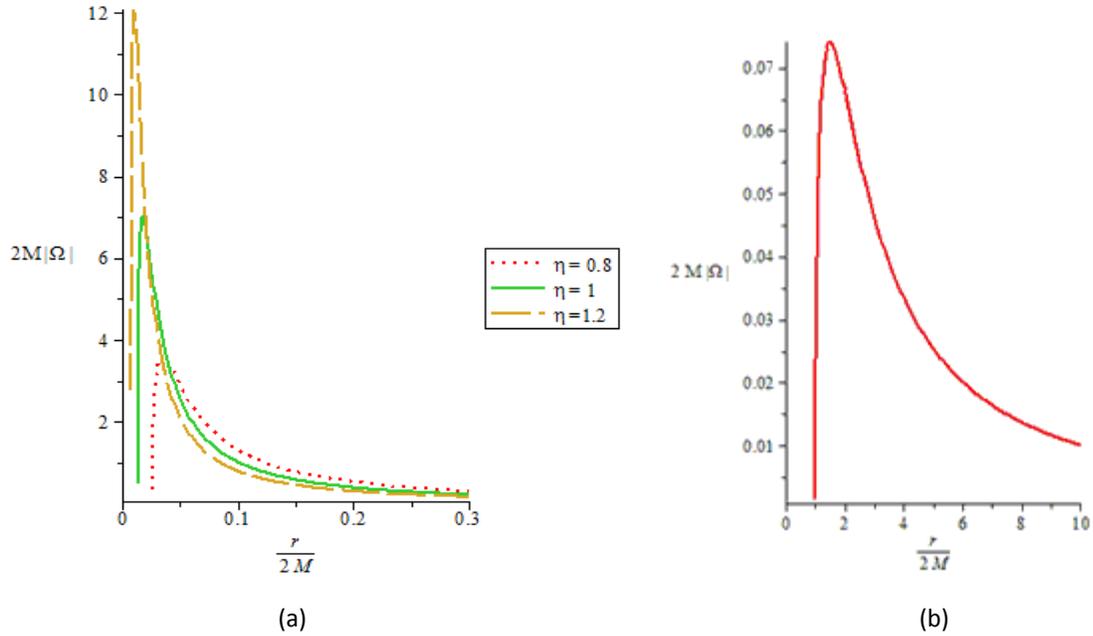

(a)                                                  (b)

Figure 2: Neutrino spin oscillation frequency versus  radius of the neutrino orbit for different values of  $\eta$ (a), and the commutative case  $X \to \infty$ (b).

.

# Phenomenological application

We consider the simple bipolar neutrino system which helps us to understand many qualitative features of collective neutrino oscillations in supernovae. The system composed of a homogeneous and isotropic gas that initially consists of mono-energetic $\nu_e$ and $\bar{\nu}_e$, and is described by the flavor pendulum [13]. We introduce $\epsilon$ as the fractional excess of neutrino over antineutrinos, $\epsilon = \frac{n_\nu}{n_{\bar{\nu}}}$. Fig. (1) shows that the period of transition probability is higher for the case of the Schwarzschild metric in non-commutative spaces:

$$T_{\theta=0} < T_{\theta \neq 0}. \tag{19}$$

So at a later time t > 0, we have:

$$\epsilon_t^{\theta=0} < \epsilon_t^{\theta \neq 0}. \tag{20}$$

This implies that the precession frequency $\Omega$ of the flavor pendulum as a function of the neutrino number density for the case of the Schwarzschild metric will be higher in commutative spaces.

We have interpreted the neutrino spin precession as neutrino-antineutrino oscillations which is true for Majorana neutrinos.

Similarly, it is seen from Fig.(2) for the RN metric that:

$$T_{\theta \neq 0} < T_{\theta=0}. \tag{21}$$

Which results in:

$$\epsilon_t^{\theta \neq 0} < \epsilon_t^{\theta=0}. \tag{22}$$

This implies that for the RN metric the precession frequency $\Omega$ of the flavor pendulum as a function of the neutrino number density will be higher in non-commutative spaces.

# Bound on Noncommutativity Parameter.

An interesting point is that by analyzing the diagrams of neutrino spin oscillations in non-commutative metrics in the previous sections, we obtain the following bound for the parameter $\eta$ :

$$\eta = \sqrt{\theta}/M > 0.1. \tag{23}$$

For instance the diagram corresponding to $\eta = 0.1$ is plotted in Fig.(3) which is not a correct physical representation of the process.

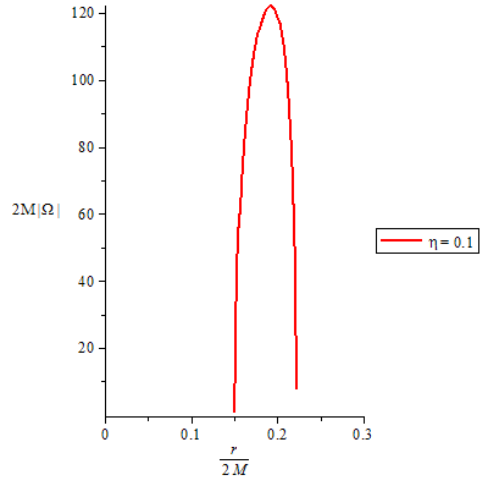

Figure 3. Neutrino spin oscillation frequency versus radius of the neutrino orbit for $\eta = 0.1$ .

As we know, the most important property of a function is that each input is related to an exactly one output. The diagram in Figure.(3) doesn't have this property. For one value of $\frac{r}{2M}$ there are more than one value for $2M|\Omega|$, specially for $2M|\Omega| < 10$ (the data shows that) it is parallel to the vertical axis. Therefore the diagram violates the definition of a function and cannot represent a function.

Using the fact that for the objects described by the metrics studied here we have $m \geq m_p$, where $m_p$ is the Planck mass, we arrive at the following bound for the noncommutativity parameter :

$$\sqrt{\theta} > 0.1 \, l_p. \tag{24}$$

which is consistent with the results reported in [7, 10, 14]. Here $l_p$ is the planck length. It is worth mentioning that an upper bound for the non-commutativity parameter was obtained in [15].

**Discussion.**

In this section we would like to discuss some interesting issues. We start with a discussion about the possible effects of non-commutativity on the geodesic equations.

In Ref. [16], authors extended the properties of non-commutativity in flat space time to curved space time. They derived the geodesic equation or in other words the force on a test particle in a noncommutative curved space time as :

$$\frac{d^2 x^i}{dt^2} = [n^i + 3G\left(\vec{\theta}^2 - \left(\vec{n}.\vec{\theta}\right)^2\right)n^i + 2G(\vec{n}.\vec{\theta})\theta^i]\frac{h}{2r}$$

where $h = \frac{-2GM}{r}$ is the Newtonian potential and $n^i = \frac{x^i}{r}$ is the radial unit vector. All the corrections are second order in $\theta$. The important point is that in this paper the authors considered the space-time non-commutativity, so their results are different from space-space non-commutativity studied by many physicists. For instance it is shown by many researchers, see e.g., [17-18] that the corrections due to space-space non-commutativity on a central force potential (including gravitational and Coulomb potentials) is first order in $\theta$ but the corrections obtained by the authors on Newtonian potential $\frac{-2GM}{r}$ are totally second order (see Eq.(5.14) of the paper). It is also shown in this paper that the time-space components of $\theta^{\mu\nu}$ are proportional to the space-space components (see, Eq.(5.13) ) :

$$\vec{\tilde{\theta}}=\pm\vec{\theta}, \text{ where } \vec{\tilde{\theta}} = \theta^{0i}, i = 1,2,3$$

So if one considers only the space-space non-commutativity i.e., $\theta^{0i} = 0$, this implies that $\vec{\tilde{\theta}}$=0, and the non-commutativity corrections on the geodesic equation vanishes. As mentioned earlier most of the papers devoted to the study of non-commutativity considered only the space-space non-commutativity. In this paper we also applied only space-space non-commutativity, so there is no corrections on the geodesic equations.

It is also interesting to check whether the neutrino spin can be directly connected to the non-commutativity parameter. In [19], authors studied the actions of the relativistic spinless and spinning particles in the framework of noncommutative field theory and extracted θ-modified actions using path integral representations of particles propagators in noncommutative field theory. They considered particles that interact with an external electromagnetic field. Their calculations show that the noncommutative corrections to the actions of the scalar (spin zero) and spinning particles are different. We would like to mention the following points :

The quantum filed theory corrections obtained in this paper are for particles which interact with real external electromagnetic field $A_\mu(x)$, with coupling $g$, but as mentioned in the text, the fields in our paper are not real external electric and magnetic fields (the coupling $g = 0$, in our case) but they are classical gravi-electric and magnetic fields.

Our study and also other works on neutrino spin oscillations use general relativity which is not a quantum field theory. But the main point is that even if we work in the framework of QFT, the corrections obtained in the mentioned paper for spin zero(scalar) and spinning particles are the same(note that in our case $g = 0$), which can be checked by comparing Eqs.(17) and (29a). Even for $g \neq 0$, the corrections are more or less the same, in particular the essential term $\dot{p}_\mu \theta^{\mu\nu} p_\nu$ is the same for spin zero and spinning particles. This means that the noncommutative corrections obtained in this paper are not due to the spin of the particles but they are coming from charge of the particles(which is zero for neutrinos) which interact with external electromagnetic field in a noncommutative background.

Now we would like to make a comment about the collective oscillations. Collective oscillations are known to happen for neutrinos that leave e.g., a supernova mainly along radial trajectories, but there is an important point involved here.

Neither in commutative nor in noncommutative spaces, there is spin oscillations for radial neutrinos in the Schwarzschild and Re-No backgrounds. As it is observed from Equations (14) and (18) of this paper the only nonzero component of the frequency i.e. $\Omega_2$ is in the $\hat{\varphi}$ direction. We get the same results for Schwarzschild and Re-No metrics in the commutative spaces by looking at Eqs.((4-9)-(4-11)) and Eq.(15) in Ref.[8] and Ref.[12], respectively. Let us present a sentence from conclusion of the Ref.[8] :

"According to Equations (4-9) - (4-11), there is no neutrino spin flip in case of radial motion in Schwarzschild metric".

For Kerr metric the situation is different and as mentioned in [12] and shown in [20], there is also spin oscillations for radial neutrinos. But at the present paper we have studied only Schwarzschild and Re-No metrics for which there is no spin oscillations for radial neutrinos. So although the fraction of circular orbit neutrinos to the radial neutrinos is small but there is no spin oscillation for radial neutrinos in Schwarzschild and Re-No backgrounds. Therefore spin oscillations is a relevant effect for trapped neutrinos, in noncommutative spaces.

The last issue we would like to address is the effects of non-commutativity on the stability of circular orbits. To study the stability of the circular orbits of particles in a central force potential (including the potential of Schwarzschild and Re-No black holes) in commutative and noncommutative spaces, we need the noncommutative effective potential. The effects of non-commutativity of space on the circular orbits of the particles in central force potential is studied in [18]. They derived the effective potential in a noncommutative background and showed that for a particle in a central force potential the noncommutative correction to the first order in $\theta$ is given by :

$$-\frac{1}{2r}(\vec{\theta}.\vec{l})\frac{\partial V}{\partial r} = -\frac{1}{2r}\theta\, l\, cos\psi\,\frac{\partial V}{\partial r}$$

where $\psi$ is the angle between $\vec{\theta}$ and $\vec{l}$.

So the non-commutativity of space manifests itself through an angular momentum dependent term. As discussed in [18], for not very large values of angular momentum the difference between commutative and noncommutative effective potentials is not considerable. Even for very large values of "$l$" for which $\cos\psi = 0$, the contribution of the non-commutativity of space on the effective potential is zero.

It is worth mentioning that even in commutative spaces not all the circular orbits with arbitrary radius are stable. For instance not all the orbits studied in [8], are stable. In this paper we also study both stable and non-stable orbits.

## Conclusion

In this paper we have studied neutrino spin oscillations in gravitational fields in non-commutative spaces. We have also analyzed the dependence of the neutrino spin oscillations frequency on the radius of the orbit. For the case of RN metric, the maximum frequency of oscillation is a monotically increasing function of the noncommutativity parameter. For the Schwarzschild metric, the maximum frequency decreases with increasing the noncommutativity parameter. We have also briefly studied the effects of noncommutativity of space on a bipolar neutrino system which reveals many qualitative features of collective neutrino oscillations in supernovae. Finally we have obtained the following bound $\sqrt{\theta} > 10^{-1} l_p$ for the noncommutativity parameter. If there exists any noncommutativity in nature, as seems to emerge from different theories and arguments, its implications should appear in neutrino oscillations in gravitational fields and in collective neutrino oscillation in supernovae.

## Appendix.

The non-zero vierbein vectors and their covariant derivatives in Schwarzschild metric are as follows :

$$e_\mu^0 = \left(\left(1 - \frac{4M}{r\sqrt{\pi}}\gamma\left(\frac{3}{2}, X\right)\right)^{\frac{1}{2}}, 0, 0, 0\right), \quad e_\mu^1 = \left(0, \left(1 - \frac{4M}{r\sqrt{\pi}}\gamma\left(\frac{3}{2}, X\right)\right)^{\frac{-1}{2}}, 0, 0\right), e_\mu^2 = (0, 0, r, 0), e_\mu^3 = (0, 0, 0, r\sin\theta), e_0^\mu =$$

$$\left(\left(1 - \frac{4M}{r\sqrt{\pi}}\gamma\left(\frac{3}{2}, X\right)\right)^{\frac{-1}{2}}, 0, 0, 0\right), \quad e_1^\mu = \left(0, \left(1 - \frac{4M}{r\sqrt{\pi}}\gamma\left(\frac{3}{2}, X\right)\right)^{\frac{1}{2}}, 0, 0\right), e_2^\mu = (0, 0, r^{-1}, 0), \quad e_3^\mu = (0, 0, 0, (r\sin\theta)^{-1}),$$

$$e_{0r;t} = \frac{\partial e_{0r}}{\partial t} - \Gamma_{tr}^t e_{0t} = \frac{1}{2}\left(1 - \frac{4M}{r\sqrt{\pi}}\gamma\left(\frac{3}{2}, X\right)\right)^{-\frac{1}{2}}\left(\frac{4M}{r^2\sqrt{\pi}}\frac{2}{3}X^{3/2} - \frac{4M}{r\sqrt{\pi}}X^{1/2}e^{-X}\right),$$

$$e_{1t;t} = \frac{1}{2}\left(1 - \frac{4M}{r\sqrt{\pi}}\gamma\left(\frac{3}{2}, X\right)\right)^{\frac{1}{2}}\left(\frac{4M}{r^2\sqrt{\pi}}\frac{2}{3}X^{3/2} - \frac{4M}{r\sqrt{\pi}}X^{1/2}e^{-X}\right),$$

$$e_{1\theta;\theta} = r\left(1 - \frac{4M}{r\sqrt{\pi}}\gamma\left(\frac{3}{2}, X\right)\right)^{\frac{1}{2}}, \qquad e_{1\varphi;\varphi} = r\left(1 - \frac{4M}{r\sqrt{\pi}}\gamma\left(\frac{3}{2}, X\right)\right)^{\frac{1}{2}} sin^2\theta,$$

$$e_{2r;r} = 1, \qquad e_{2\varphi;\varphi} = -r\sin\vartheta\cos\theta,$$

$$e_{3r;\varphi} = \sin\vartheta, \qquad e_{3\theta;\varphi} = r\cos\vartheta.$$

where $X = \frac{r^2}{4\theta}$

The non-zero vierbein vectors and their covariant derivatives in Re-No metric are given by the following expressions :

$$e_t^0 = \left(1 - \frac{4M}{r\sqrt{\pi}}\gamma\left(\frac{3}{2}, X\right) + \frac{Q^2}{\pi r^2}\gamma^2\left(\frac{1}{2}, X\right) - \frac{Q^2}{\pi r\sqrt{2\theta}}\gamma\left(\frac{1}{2}, \frac{r^2}{2\theta}\right)\right)^{\frac{1}{2}}, \qquad e_r^1 = \left(1 - \frac{4M}{r\sqrt{\pi}}\gamma\left(\frac{3}{2}, X\right) + \frac{Q^2}{\pi r^2}\gamma^2\left(\frac{1}{2}, X\right) - \frac{Q^2}{\pi r\sqrt{2\theta}}\gamma\left(\frac{1}{2}, \frac{r^2}{2\theta}\right)\right)^{-\frac{1}{2}},$$

$$e_\theta^2 = r, \; e_\varphi^3 = r\sin\theta.$$

$$e_{0r;t} = -\frac{1}{2}\left[\frac{4M}{r^2\sqrt{\pi}}\gamma\left(\frac{3}{2}, \frac{r^2}{4\theta}\right) - \frac{4M}{r\sqrt{\pi}}\frac{d}{dr}\gamma\left(\frac{3}{2}, \frac{r^2}{4\theta}\right) - \frac{2Q^2}{\pi r^3}\gamma\left(\frac{3}{2}, \frac{r^2}{4\theta}\right)^2 + \frac{Q^2}{\pi r^2}\frac{d}{dr}\gamma^2\left(\frac{3}{2}, \frac{r^2}{4\theta}\right) + \frac{Q^2}{\pi r^2\sqrt{2\theta}}\gamma\left(\frac{1}{2}, \frac{r^2}{2\theta}\right)\right.$$

$$\left. - \frac{Q^2}{\pi r\sqrt{2\theta}}\frac{d}{dr}\gamma\left(\frac{1}{2}, \frac{r^2}{2\theta}\right)\right]\left[1 - \frac{4M}{r\sqrt{\pi}}\gamma\left(\frac{3}{2}, \frac{r^2}{4\theta}\right) + \frac{Q^2}{\pi r^2}\gamma\left(\frac{3}{2}, \frac{r^2}{4\theta}\right)^2 - \frac{Q^2}{\pi r\sqrt{2\theta}}\gamma\left(\frac{1}{2}, \frac{r^2}{2\theta}\right)\right]^{-\frac{1}{2}},$$

$$e_{1t;t} = \frac{1}{2}\left[\frac{8M}{3r^2\sqrt{\pi}}X^{\frac{3}{2}} + \frac{4M}{r\sqrt{\pi}}X^{\frac{1}{2}}e^{-X} - \frac{8Q^2X}{\pi r^3} + \frac{4Q^2}{\pi r^2}e^{-X} - \frac{2Q^2X^{\frac{1}{2}}}{\pi r^2\sqrt{2\theta}} + \frac{Q^2X^{-\frac{1}{2}}e^{-X}}{\pi r\sqrt{2\theta}}\right]\left[1 - \frac{8MX^{\frac{3}{2}}}{r\sqrt{\pi}} + \frac{4Q^2X}{\pi r^2} - \frac{2Q^2X^{\frac{1}{2}}}{\pi r\sqrt{2\theta}}\right]^{\frac{1}{2}},$$

$$e_{1\theta;\theta} = r\left[1 - \frac{8MX^{\frac{3}{2}}}{r\sqrt{\pi}} + \frac{4Q^2X}{\pi r^2} - \frac{2Q^2X^{\frac{1}{2}}}{\pi r\sqrt{2\theta}}\right], \quad e_{1\varphi;\varphi} = r\sin^2\vartheta\left[1 - \frac{8MX^{\frac{3}{2}}}{r\sqrt{\pi}} + \frac{4Q^2X}{\pi r^2} - \frac{2Q^2X^{\frac{1}{2}}}{\pi r\sqrt{2\theta}}\right],$$

$$e_{2r;\theta} = 1, \qquad e_{2\varphi;\varphi} = -r\sin\vartheta\cos\vartheta,$$

$$e_{3r;\varphi} = \sin\vartheta, \qquad e_{3\theta;\varphi} = r\cos\vartheta.$$

**Acknowledgement.**

S. A. Alavi would like to thank the Department of Physics of the University of Torino for hospitality where some parts of this work were done. He also very grateful to Samoil Bilenkey (JINR Dubna) and Carlo Giunti (INFN, Torino) for useful discussions.